\documentclass{lmsedition}
\usepackage{enumitem}
\usepackage[super]{nth}
\usepackage[none]{hyphenat}
\setlist[itemize]{leftmargin=*}
\setlist[enumerate]{leftmargin=*}

  {\list{}{\leftmargin=5mm \rightmargin=5mm}\item[]}%
  {\endlist}

\makeatletter
\def\fakeH@#1#2{%
  \vtop{\m@th\ialign{##\cr
    \hfil$#1\kern-.8pt\operator@font \textsf{o}$\hfil\cr
    \noalign{\nointerlineskip\kern-7\ex@}#2\cr
    \noalign{\nointerlineskip\kern-\ex@}\cr}}%
}
\def\fakHo{%
  \mathop{\mathpalette\fakeH@{\'{}\kern-.6pt\'{}}}\nmlimits@
\!}
\makeatother

\makeatletter
\def\fakec@#1#2{%
  \vtop{\m@th\ialign{##\cr
    \hfil$#1\kern-.8pt\operator@font \textsf{c}$\hfil\cr
    \noalign{\nointerlineskip\kern-7\ex@}#2\cr
    \noalign{\nointerlineskip\kern-\ex@}\cr}}%
}
\def\fakdc{%
  \mathop{\mathpalette\fakec@{\'{}}}\nmlimits@
\!}
\makeatother

\newcommand{\Z}{\mathbb{Z}}

\newcommand{\F}{\mathbb{F}}

\begin{document}

\articletitle{Girth of the Cayley graph and Cayley hash functions}

\articleauthor{Vladimir Shpilrain}

%

 
\begin{articlenoabs}
Cayley hash functions are based on a simple idea of using a pair of
semigroup elements,  $A$ and  $B$, to hash the 0 and 1 bit,
respectively, and  then to hash an arbitrary bit string in the
natural way, by using multiplication of elements in the semigroup. The main advantage of Cayley hash functions compared to, say, hash functions in the SHA family is that when an already hashed document is amended, one does not have to hash the whole amended document all over again, but rather hash just the amended part and then multiply the result by the hash of the original document. In this article, we survey some of the previously proposed Cayley hash functions and single out a very simple hash function whose security has not been compromised up to date. 
\end{articlenoabs}

\begin{twoblock}
 

\section*{Introduction}
 
Hash functions are easy-to-compute compression functions that take
a variable-length input and convert it to a fixed-length output.
Hash functions are used as compact representations, or digital
fingerprints, of data and to provide message integrity. {\it Cryptographic hash functions} have many information-security applications, notably in digital signatures, message authentication codes, and other forms of authentication. Cryptographic hash functions should satisfy the following basic security requirements:

\begin{enumerate}

\item {\it Collision resistance}: it should
be computationally infeasible to find two different inputs that hash
to the same output.

\item {\it Preimage resistance} (sometimes called {\it non-invertibility}): it should be
computationally infeasible to find an input which hashes to a specified output.

\item {\it Second preimage resistance}: it
should be computationally infeasible to find a second input that
hashes to the same output as a specified input.
 
\end{enumerate}

A challenging problem is to determine mathematical properties of a
hash function that would ensure (or at least, make it likely) that
the requirements above are met.
 
 A direction that has been gaining momentum lately is using a pair of
elements, $A$ and $B$, of a semigroup $S$, to hash the ``0" and the ``1" bit, respectively. Then a bit string is hashed to a product of elements in the natural way.
For example, the bit string 1001011 will be hashed to  the element $BAABABB$.

Since hashing a random bit string this way represents a random walk on the Cayley
graph of the subsemigroup of $S$ generated by the elements $A$ and $B$, hash functions of this kind are often called {\it Cayley hash functions}. Note that the absence of short collisions for a Cayley hash function is equivalent to the corresponding Cayley graph having a large {\it girth}. The latter is defined as the length of the shortest simple circuit.
 
Cayley hash functions have a homomorphic property $H(XY)=H(X)H(Y)$ and the
associativity property $H(XYZ)=H(XY)H(Z) = H(X)H(YZ)$ for any bit
strings $X, Y, Z$. (Here $XY$ means concatenation of the bit strings $X$
and $Y$.) This property is useful not only because it allows for parallel computations  when hashing a long bit string. A more important feature is: when an already hashed document is amended, one does not have to hash the whole amended document all over again, but rather hash just the amended part and then multiply the result by the hash of the original document. 
 
Another useful property of a Cayley hash function is that, unlike with some other hash functions, you do not have to know the length of a bit string to be hashed up front; you can hash ``as you go".

While the high-level idea of Cayley hashing is definitely appealing, the choice of the platform semigroup $S$ and two elements $A, B \in S$ is crucial for security and efficiency. There have been many proposals based on matrix semigroups in $GL_2(\F)$ for various fields $\F$, in particular for $\F=\F_p$. This is because Cayley graphs of 2-generator semigroups in $GL_2(\F_p)$ often have a large girth as was shown by several authors, see e.g. \cite{BG}, \cite{BSV}, \cite{Masuda}, \cite{H}. 

Cayley graphs of (semi)groups in $GL_n(\F_p)$ with $n >2$ have been considered, too (see e. g.  \cite{Battarbee}), but we will focus here on $n =2$, one of the reasons being a smaller size of the hash. For example, if $p$ is a 256-bit prime, then any matrix from $GL_2(\F_p)$ has total  size of up to 1024 bits, which is common for standard hash functions these days.

\section*{Specific platform (semi)groups}\label{previous}

The first proposal of a Cayley hash function was due to Z\'emor \cite{Zemor}.
The matrices used, considered over $\F_p$, were 
$A = \left(
 \begin{array}{cc} 1 & 1 \\ 0 & 1 \end{array} \right) , \hskip 1cm B = \left(
 \begin{array}{cc} 1 & 0 \\ 1 & 1 \end{array} \right).$

This proposal was successfully attacked in \cite{TZ_attack}. Specifically, it was shown that this hash function is not preimage resistant.

\begin{wrapfigure}{l}{32mm}
\includegraphics[width=32mm]{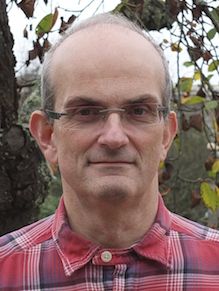}
\caption*{\textcolor{dcolor}{Figure 1. Gilles Zémor, Université de Bordeaux}} 
\end{wrapfigure}

The most cited proposal is what has become known as the Tillich-Z\'emor hash function \cite{TZ}. Their matrices were 
$A = \left(
 \begin{array}{cc} \alpha & 1 \\ 1 & 0 \end{array} \right) , \hskip 1cm B = \left(
 \begin{array}{cc} \alpha & \alpha+1 \\ 1 & 1 \end{array} \right).$

\noindent  These matrices are considered over a field defined as
$R=\F_2[x]/(p(x))$, where  $\F_2[x]$ is the ring of polynomials over
$\F_2$,  $(p(x))$ is the ideal of $\F_2[x]$
generated by an irreducible polynomial $p(x)$ of degree $n$
(typically,  $n$ is a prime, $127 \le n \le 170$), and  $\alpha$ is a root of $p(x)$.

The reason for selecting such a ``fancy" field probably was to specifically avoid the attack in \cite{TZ_attack}.

\begin{wrapfigure}{l}{32mm}
\includegraphics[width=32mm]{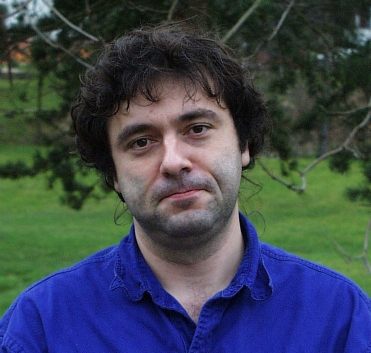}
\caption*{\textcolor{dcolor}{Figure 2. Jean-Pierre Tillich, Centre Inria de Paris}} 
\end{wrapfigure}

Similar later proposals and attacks (some of them targeted at finding collisions, some targeted at finding a preimage) were suggested over the years, see e.g. \cite{cookies} for a list of relevant references.  

A simple yet fruitful idea for avoiding short collisions is to use a pair of $2 \times
2$ matrices, $A$ and $B$, over $\Z$  that generate a free semigroup in $GL_2(\Z)$,
and then reduce the entries modulo a large prime $p$ to get matrices
over $\F_p$. Since there cannot be an equality of two different
products of copies of $A$ and $B$ unless at least one of
the entries in at least one of the products is $\ge p$, this gives a
lower bound on the minimum length of bit strings where a collision
may occur.

\section*{Girth of the Cayley graph}\label{growth}

The problem of bounding the girth of the Cayley graph of a 2-generator (semi)group is directly related to security properties (specifically, to collision resistance) of the relevant Cayley hash functions. 

In the case of matrix semigroups, if $A$ and $B$ generate a free sub(semi)group of $SL_2(\Z)$, then there cannot be any relations of the form $u(A, B)=v(A, B)$ in $SL_2(\Z_p)$ unless at least one of the entries of the matrix $u(A, B)$ or $v(A, B)$ is at least $p$. Thus, if the largest entry in a product of $n$ matrices is of the size $O(s^n)$, then the girth of the Cayley graph of the  sub(semi)group of $SL_2(\Z_p)$ generated by $A$ and $B$ is $O(\log_s p)$.
This (maximal) growth rate $s$ is called the {\it joint spectral radius} of the pair $(A, B)$ of matrices and has been studied (in greater generality) a lot, see e.g. \cite{Jungers}.

We are interested in having the joint spectral radius of $(A, B)$ as small as possible to have a larger girth of the corresponding Cayley graph. To that end, let us consider pairs of matrices $(A(k), B(m))$, where $A(k) = \left(
 \begin{array}{cc} 1 & k \\ 0 & 1 \end{array} \right) , \hskip .2cm B(m) = \left(
 \begin{array}{cc} 1 & 0 \\ m & 1 \end{array} \right)$. The Cayley graph of the (semi)group generated by these two matrices (considered over $\Z_p$), especially when $m=k$, has been extensively studied in the literature, see e.g. \cite{BG}, \cite{H} and references therein. 

When considered over $\Z$, the joint spectral radius of the pair of matrices $(A(k), B(k))$ for $k \ge 1$ was computed, in particular, in \cite{BSV} and \cite{Masuda}. In fact, \cite{BSV} gives explicit formulas for the largest entry in a product of $n$ copies of $A(k)$ and $B(k)$. 
As expected, the smallest joint spectral radius is achieved when $k=1$ (and is equal to $\frac{1}{2}(1+\sqrt{5}) \approx 1.618$), but as we mentioned before, the corresponding Cayley hash function was successfully attacked in \cite{TZ_attack}. 

Therefore, \cite{BSV} proposed using the Cayley hash function corresponding to the pair of matrices $(A(2), B(2))$, where the joint spectral radius is $1+\sqrt{2} \approx 2.414$. 
It is worth mentioning that powers of the matrix $A(2)B(2)$ provide the largest (by the absolute value) entries among all semigroup words in $A(2)$ and $B(2)$ of a given length.

This implies, in particular, that if $p$ is, say, on the order of $2^{256}$,  then there
are no collisions of the form  $u(A(2), B(2)) = v(A(2), B(2))$ if both the
words $u$ and  $v$ are of length less than $201 \approx  256 \log_{2.414}2$. This makes a ``brute force" search for collisions computationally infeasible.

We note that up to date, there have been no successful attacks reported against the Cayley hash function based on the matrices $A(2)$ and $B(2)$. It is also worth mentioning that this hash function has successfully passed all the
pseudorandomness tests in the NIST Statistical Test Suite \cite{NIST}.

\section*{Cayley hashing with cookies}\label{cookies}

In \cite{cookies}, the authors introduced an enhancement of Cayley hashing that they called ``Cayley hashing with cookies", the terminology borrowed from the theory of random walks in a random environment. The authors argue that this enhancement does not affect the collision resistance property, and at the same time makes the hash function more preimage resistant. The homomorphic property is ``almost preserved", i.e., is preserved upon minor padding.

A ``cookie" is a place in the Cayley graph where some of the parameters of a random walk change in a specific way. There is a lot of flexibility in positioning cookies in the Cayley graph as well as in choosing a particular way a cookie affects parameters of a random walk on the Cayley graph. Below is an example (from \cite{cookies}) of an instantiation of this general idea.

Let $A$, $B$, and $C$ be $2 \times 2$ matrices. Let $u$ be a bit string of an arbitrary length. Then, to hash $u$, going left to right:

\medskip

\noindent {\bf 1.} If the current bit is 0, then it is hashed to the matrix $A$.
If the current bit is 1, then it is hashed to the matrix $B$. 

\medskip

\noindent {\bf 2.} If there are three ``1" bits in a row (a ``cookie"), then all following ``1" bits will be hashed to the matrix $C$, until there are three ``0" bits in a row, in which case hashing the ``1" bit is switched back to the matrix $B$.
For example, the bit string 1100111 01011 00011 will be hashed to the matrix $BBAABBB ACACC AAABB$.

\noindent In \cite{cookies}, the recommended particular matrices were: $A = \left(
 \begin{array}{cc} 1 & 2 \\ 0 & 1 \end{array} \right) , \hskip .2cm B = \left(
 \begin{array}{cc} 1 & 0 \\ 2 & 1 \end{array} \right), \hskip .2cm C = \left(
 \begin{array}{cc} 2 & 1 \\ 1 & 1 \end{array} \right)$.

These matrices generate a free semigroup when considered over $\Z$. It was shown in \cite{cookies} that the joint spectral radius of the triple of matrices of $(A, B, C)$ as above is $\frac{7}{2}+\frac{3\sqrt{5}}{2} \approx 2.618$.

\section*{Directions of further research}\label{Directions}

Most of the theoretical results (if not all of them) on the joint spectral radius of matrices, see e.g. \cite{Jungers}, are relevant to matrices with nonnegative entries. However, having in mind our goal of minimizing the joint spectral radius, there is nothing wrong with using matrices some of whose entries are negative. 

An obvious candidate here would be the pair $(A(2), B(-2))$, where $A(2) = \left(
 \begin{array}{cc} 1 & 2 \\ 0 & 1 \end{array} \right) , \hskip 1cm B(-2) = \left(
 \begin{array}{cc} 1 & 0 \\ -2 & 1 \end{array} \right).$

We do not know what the joint spectral radius of this pair of matrices is, but computer experiments suggest that it is $\sqrt{2+\sqrt{3}} \approx 1.93$, so it is considerably smaller than $1+\sqrt{2} \approx 2.414$, the joint spectral radius of the pair $(A(2), B(2))$.
It would be good though to establish this result theoretically. We also note that, again based on computer experiments, it appears that powers of the matrix $A(2)^2B(-2)^2$ provide the largest (by the absolute value) entries among all semigroup words in $A(2)$ and $B(-2)$ of a given length.

Another direction of research related to Cayley hashing (with matrices $A$ and $B$) is motivated by the fact that since bit strings that are hashed in real-life applications can be considered random, one might look at the length of ``generic"  simple circuits of the relevant Cayley graph instead of looking for the length of the shortest simple circuit. To that end, one can consider products of $n$ matrices, where each factor is either $A$ or $B$ with probability $\frac{1}{2}$ and see how the largest entry in such a product grows when $n$ goes to infinity. This yields interesting connections to the theory of stochastic processes, see e.g. \cite{Pollicott}.

\begin{wrapfigure}{l}{32mm}
\includegraphics[height=36mm, width=32mm]{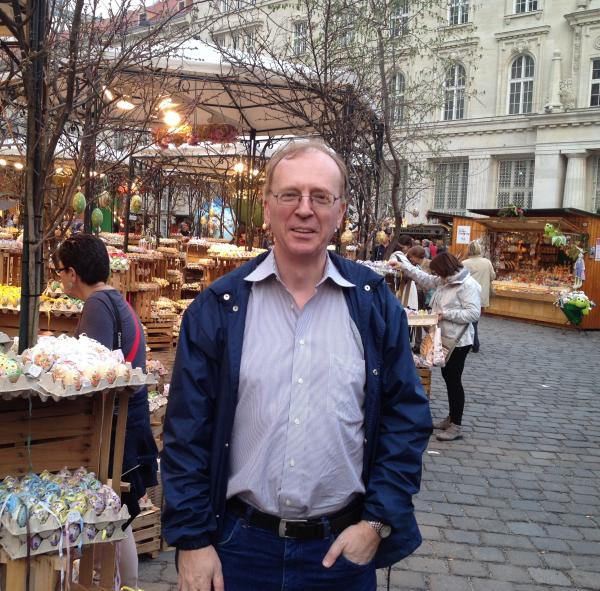}
\end{wrapfigure} 
\textbf{{\large Vladimir Shpilrain}}  \\~\\
Vladimir is a professor of mathematics at the City College of New York and a member of the doctoral faculty at the CUNY Graduate Center.  His main research interests are in information security, complexity of algorithms, combinatorial and computational group theory.


\end{twoblock}
 

\begin{thebibliography}{ABC}
 
\bibitem{BG}
J. Bourgain, A. Gamburd, {\it Uniform expansion bounds for Cayley
graphs of $SL_2(\F_p)$}. Ann. of Math. (2) {\bf 167}
(2008),  625--642.

\bibitem{BSV}
L. Bromberg, V. Shpilrain, A. Vdovina, {\it
Navigating in the Cayley graph of $SL_2(\F_p)$ and applications to
hashing}, Semigroup Forum {\bf 94} (2017), 314--324.
 

\bibitem{Masuda}
S. Han, A. M. Masuda, S. Singh, J. Thiel, {\it Maximal entries of elements
in certain matrix monoids}, Integers {\bf 20} (2020), paper No. A31.

\bibitem{H}
H. A. Helfgott,{\it Growth and generation in $SL_2(\Z/p\Z$)} Ann. of
Math. (2) {\bf 167} (2008),  601--623.

\bibitem{Jungers}
R. Jungers, {\sl The Joint Spectral Radius: Theory and Applications}, Springer Lecture Notes in Control and Information Sciences {\bf 385}  (2009), 160 pp.


\bibitem{Battarbee}
C. Le Coz, C. Battarbee, R. Flores, T. Koberda, D. Kahrobaei, {\it Post-quantum
hash functions using $SL_n(\F_p)$}, \url{https://arxiv.org/abs/2207.03987}

\bibitem{NIST}
National Institute of Standards and Technology - NIST , \emph{NIST
Statistical Test Suite}, 2010. \url{http://csrc.nist.gov/groups/ST/toolkit/rng/documentation\_software.html}


\bibitem{Pollicott}
M. Pollicott, {\it Maximal Lyapunov exponents for random matrix products}, Invent. Math. {\bf 181} (2010), 209--226.

\bibitem{cookies}
V. Shpilrain, B. Sosnovski, {\it Cayley hashing with cookies}, in: Future of Information and Communication Conference (FICC 2025), Springer Lecture Notes in Networks and Systems, to appear.  \url{https://arxiv.org/abs/2402.04943}

\bibitem{TZ_attack}
J.-P. Tillich and G. Z\'emor,  {\it Group-theoretic hash functions},
in Proceedings of the First French-Israeli Workshop on Algebraic
Coding, Lecture notes  Comp. Sci. {\bf  781} (1994),   90--110.

\bibitem{TZ}
J.-P. Tillich and G. Z\'emor, {\it Hashing with $SL_2$}, in CRYPTO
1994, Lecture Notes  Comp. Sci. {\bf 839} (1994), 40--49.

\bibitem{Zemor}
G. Z\'emor, {\it  Hash Functions And Graphs With Large Girths  in Eurocrypt'91}, Lecture Notes in Comput. Sci. {\bf 547} (1991), 508--511.


\end{thebibliography}
\end{document}